\documentclass[reprint,article,showpacs,preprintnumbers,amsmath,amssymb,citeautoscript,aip,superscriptaddress]{revtex4-1}
\usepackage{graphicx,wasysym}
\usepackage{dcolumn}
\usepackage{bm}
\usepackage{hyperref}
\usepackage{color}
\usepackage[normalem]{ulem}

\setcitestyle{super}

\newcolumntype{P}[1]{>{\centering\arraybackslash}p{#1}}

\makeatletter

\newcommand{\thickhline}{%
    \noalign {\ifnum 0=`}\fi \hrule height 0.4pt
    \futurelet \reserved@a \@xhline}
\newcolumntype{"}{@{\hskip\tabcolsep\vrule width 1pt\hskip\tabcolsep}}
\makeatother

\begin{document}

\preprint{1}

\begin{abstract}

Abstract: Polariton chemistry holds promise for facilitating mode-selective chemical reactions, but the underlying mechanism behind the rate modifications observed under vibrational strong coupling is not well understood.
Using the recently developed quantum transition path theory, we have uncovered a mechanism of resonant suppression of a thermal reaction rate in a simple model polaritonic system, consisting of a reactive mode in a bath confined to a lossless microcavity with a single photon mode. We observed the formation of a polariton during rate limiting transitions on reactive pathways and identified the concomitant rate suppression as due to hybridization between the reactive mode and the cavity mode, which inhibits bath-mediated tunneling. 
The transition probabilities that define the quantum master equation can be directly translated into a visualisation of the corresponding polariton energy landscape. This landscape exhibits a double funnel structure, with a large barrier between the initial and final states.

\end{abstract}

\title{On the Mechanism of Polaritonic Rate Suppression from \\ Quantum Transition Paths}

\author{Michelle C. Anderson}
\affiliation{
Department of Chemistry, University of California, Berkeley 94720, USA \looseness=-1
}
\author{Esmae J. Woods}
\affiliation{Department of Physics, University of Cambridge, Cambridge CB3 0HE, UK \looseness=-1}
\affiliation{Yusuf Hamied Department of Chemistry, University of Cambridge, Cambridge CB2 1EW, UK \looseness=-1}
\author{Thomas P. Fay}
\affiliation{
Department of Chemistry, University of California, Berkeley 94720, USA \looseness=-1
}
\author{David J. Wales}
\affiliation{Yusuf Hamied Department of Chemistry, University of Cambridge, Cambridge CB2 1EW, UK \looseness=-1}
\author{David T. Limmer} \email{dlimmer@berkeley.edu}
\affiliation{
Department of Chemistry, University of California, Berkeley 94720, USA \looseness=-1
}
\affiliation{
Kavli Energy NanoSciences Institute, University of California, Berkeley 94720, USA \looseness=-1
}\affiliation{
Chemical Sciences Division, Lawrence Berkeley National Laboratory, Berkeley 94720, USA \looseness=-1
}
\affiliation{%
Materials Sciences Division, Lawrence Berkeley National Laboratory, Berkeley 94720, USA \looseness=-1
}

\date{July 11, 2023}


\maketitle

When molecules are confined to a dark microcavity a vibrational polariton can form from the hybridization between a molecular vibrational mode and the vacuum photon mode of the cavity. The rates of ground state bond breaking reactions of molecules in such cavities have been shown to be significantly modulated from their values outside of the cavity. \cite{thomas_lethuillier-karl,thomas_george_shalabney,lather_bhatt_thomas} This phenomenon of polaritonic chemistry holds promise for selective catalysis, however the mechanisms of rate enhancement or suppression remain unclear. Here, we have applied quantum transition path theory (QTPT)\cite{anderson_schile_limmer} to a simple model in order to elucidate the underlying quantum mechanical source of the rate modification. We find that the origin of sharp rate decreases under cavity resonance conditions is due to poor bath-mediated tunneling between polariton wavefunctions.

In experimental and theoretical studies, polariton formation has been shown to  affect reactive behavior in both the excited and ground state.\cite{ribeiro_martinez,mandal_huo,pino_feist,hutchinson_schwartz,hoffmann_lacombe,thomas_lethuillier-karl,lather_bhatt_thomas} Of particular interest is the change in rates observed in ground state barrier crossing reactions, when molecules are confined to a microcavity with a vacuum mode in resonance with key molecular vibrations.\cite{lather_bhatt_thomas,thomas_lethuillier-karl,thomas_george_shalabney} Attempts to explain the experimentally observed sharp changes in behavior under resonance conditions have met with mixed success and different interpretations. Classical transition state theories do not support resonance effects.\cite{cga_jorge_yz,zhadanov,li_nitzan_subotnik,galego_climent} Other work has used extensions to transition state theory \cite{grote_hynes,lindoy_lachlan,voth_chandler,jianshu_voth} to explain the rate suppression on resonance in terms of dynamical caging effects\cite{li_mandal} or tunneling effects.\cite{voth_chandler,jianshu_voth} Although these studies have reproduced rate modifications, they have tended to reveal broad, shallow rate suppression  under resonance, which does not  agree with experimental observations that indicate sharp rate modifications.\cite{li_mandal,li_nitzan,lindoy_nature} A full quantum dynamical study using the hierarchical equations of motion\cite{tanimura_kubo} carried out by Lindoy and coworkers has recently identified sharp rate modifications under resonance conditions\cite{lindoy_nature}. This result lends credence to the notion that intrinsically quantum mechanical effects must be modeled in order to observe polaritonic rate enhancement or suppression. \cite{lindoy_nature,li_nitzan_subotnik} 

To elucidate the source of resonant effects in polaritonic systems, we have employed a simple Pauli-Fierz\cite{pauli_fierz} quantum electrodynamics Hamiltonian\cite{li_mandal,flick_ruggenthaler} for a single photon mode coupled to a reactive proton coordinate solvated in a bath, and employed QTPT to extract barrier crossing rates and mechanisms. QTPT and related quantum path sampling techniques have been recently developed and used to extract mechanistic information from quantum dynamical processes, including energy transfer\cite{schile_limmer,arsenault2021vibronic} and nonadiabatic relaxation through conical intersections.\cite{anderson_schile_limmer,schile2019simulating} 
Here we have used QTPT to extract the dominant reactive pathways of a thermally induced proton transfer event under conditions where the proton was resonantly coupled to a cavity photon mode, whose natural frequency we could adjust. These pathways are given by a series of jumps through energy eigenstates of the combined proton-cavity system. After analyzing this dominant reactive pathway to determine the committor eigenstates, which correlate with the classical transition state, we find the fall in rate is caused by reduced tunneling matrix elements between the committor eigenstates, caused by the formation of polaritons. Our key result is that the poor overlap is due to the formation of polaritonic wavefunctions under resonance conditions.

We address a model similar to the Shin-Metiu formulation\cite{shin_metiu} employed in several previous studies,\cite{li_mandal,li_nitzan} under the Pauli-Fierz Hamiltonian\cite{pauli_fierz,du_poh} in which light and matter are treated quantum mechanically. In this model, the long wavelength approximation in the dipole gauge is followed by the Power-Zienau-Wooley transformation. The resulting Hamiltonian includes a dipole self-energy term which, if neglected, will result in an incorrect potential.\cite{cohen-tannoudji,power_zienau} Since it is convenient for QTPT to have localized eigenstate wavefunctions on either side of a barrier to define the reactant and product states, we added a small linear bias to remove bistability of the original  Shin-Metiu model. Due to the relatively high mass of the proton coordinate, the bias magnitude necessary to localize wavefunctions on either side of the barrier was very small. The resulting system Hamiltonian, $H_s$ is,
\begin{equation}
\begin{split}
H_s &= P^2/(2M) + U(R) + p_c^2/2  \\ & 
+ \frac{\omega_c^2}{2} \left(q_c + \sqrt{2/(\hbar \omega_c^3}) \chi \mu(R)\right)^2 ,
\end{split}
\end{equation}
where $P$ and $R$ are the proton momentum and position, $p_c$ and $q_c$ are the corresponding photon coordinates, $\omega_c$ is the photon frequency, $\hbar$ is Plank's constant, $\mu(R)$ is the proton dipole operator, $U(R)$ is the potential energy of the proton coordinate, and $\chi$ is a parameter which controls the coupling strength between light and matter. 
The coupling of the cavity to the system dipole should be interpreted as being dependent on the number of reactive molecules in the cavity, which under the assumption that the dipolar molecules' motion is independent and isotropic, can be decoupled.\cite{du_poh}
The resultant Born-Oppenheimer surface is given by $E(R,q_c) = H_s - P^2/(2M) - p_c^2/2$. Note we do not consider the effects of cavity leakage here, which has been shown to be important in certain regimes.\cite{mandal_li_huo_2022,lindoy_nature}

\begin{figure}[t]
\begin{center}
\includegraphics[width=8.5cm]{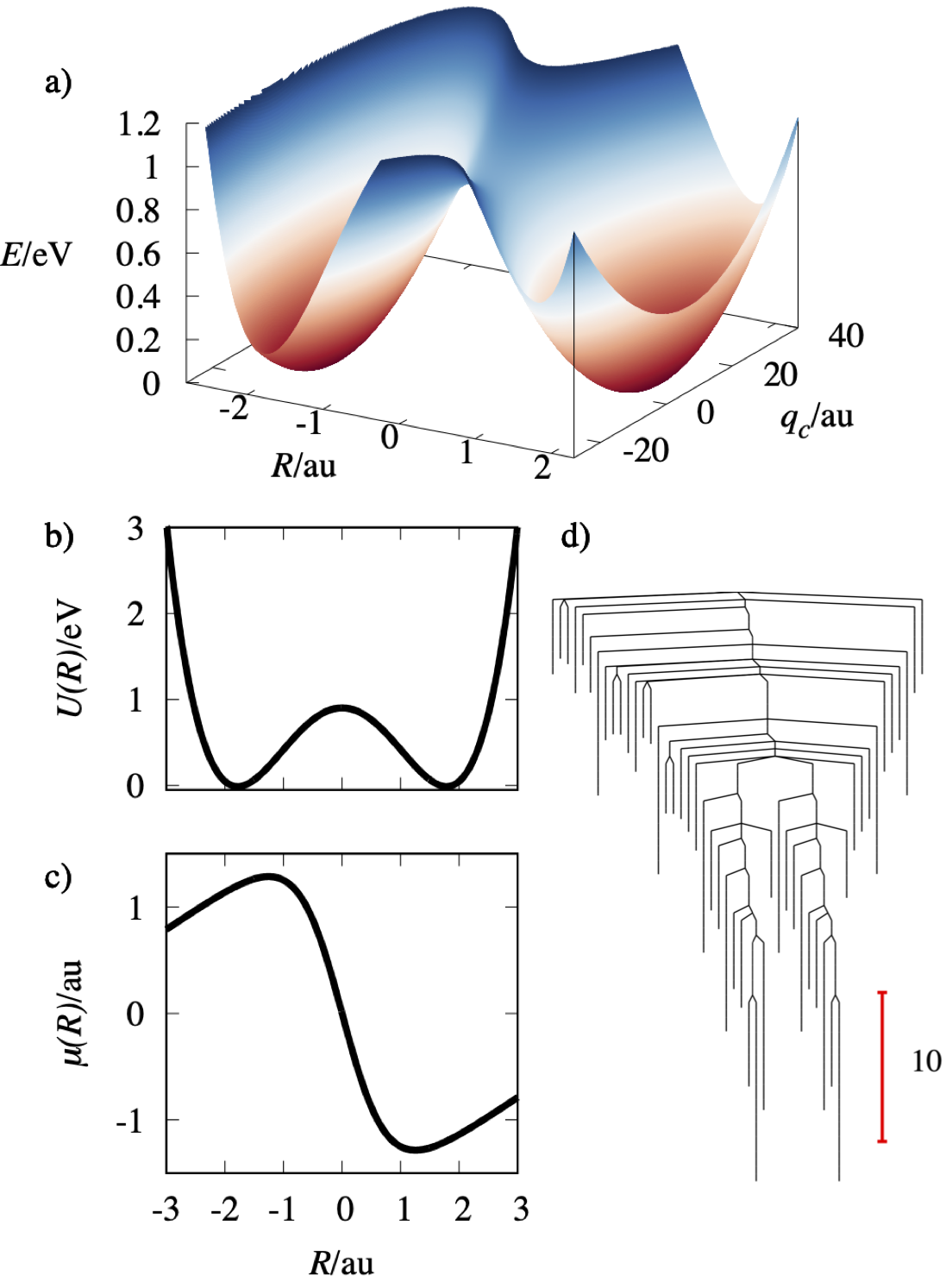}
\caption{a) Potential energy surface of the proton-photon system when $\omega_c = 0.925\omega_s$ and $\eta_c=\eta_s$ and a polariton is expected to form. b) The potential energy surface of the proton coordinate. c) The proton dipole operator. d) Free energy disconnectivity graph\cite{beckerk97,WalesMW98} corresponding to the quantum master equation that governs the dynamics. The resonant states lie at the bottom of distinct funnels separated by a high effective barrier. The red line indicates a free energy scale of 10 $k_BT$.}\label{pes}
\end{center} 
\end{figure}

The functions used for $E(R,q_c)$, $U(R)$ and $\mu(R)$ are illustrated in Fig. \ref{pes}. These potentials are similar to those employed by Li and coworkers,\cite{li_mandal} with explicit forms given in the supporting information (SI). The potential energy surface, $E(R,q_c)$ is shown in Fig. \ref{pes} a) for the case where the system is in resonance and polariton formation occurs. Note that the bottoms of the wells illustrated in $E(R,q_c)$ are not centered at $q_c=0$ but displaced to either side, whereas the surface remains effectively symmetric about 0 in $R$. The proton potential in Fig. \ref{pes} b) is a simple double well reflecting two distinct covalently bonded states of the proton. The form of the position dependent dipole shown in Fig. \ref{pes} c) is consistent with the notion that a positively charged proton is moving between the two metastable states.  The resultant eigenstates are more clearly shown by Fig. \ref{pes} d) which displays a free energy disconnectivity graph\cite{beckerk97,WalesMW98}  for the quantum master equation corresponding to an on resonance system. The symmetric bifurcation corresponds to bistability, where the resonant states lie at the bottom of distinct funnels separated by a high barrier.  

To address the dynamics of the system, QTPT was applied to an open system description with full Hamiltonian,\cite{nitzan,breuer_petruccione} 
\begin{equation}
    H = H_s + H_B + R \otimes B ,
\end{equation}
in which the total Hamiltonian is broken down into $H_s$ which operates only on the system, $H_B$, which operates only on the bath, and a coupling operator $R \otimes B$. The bath is envisioned to include all non-reactive modes of the system, including molecular and solvent modes. Additionally, the bath captures interactions between the reactive mode of the molecule and other reactive molecules whose dipoles are aligned with the cavity.\cite{du_poh}
The bath is approximated by an infinite set of harmonic oscillators that relax quickly in comparison to the system dynamics, which allows the bath effects to be addressed perturbatively via the Born-Markov approximation.\cite{redfield,nitzan,breuer_petruccione} The coupling operator involves $R$, the position operator of the proton, via the tensor product with $B= \sum_k c_k R_k$, a sum over position coordinates, $R_k$, of  the bath harmonic oscillators, with coupling strength parameters, $c_k$ determined by the spectral density,
 \begin{equation}
     J(\omega) = \frac{\pi}{2} \sum_k \frac{c_k^2}{\omega_k} \delta(\omega - \omega_k) = \eta \omega e^{-|\omega|/\omega_b},
\end{equation}
in which  $\omega_k$ is the frequency of bath oscillator $k$, $\omega_b$ is the bath cut-off frequency, and $\eta$ is the system-bath coupling strength. To employ QTPT, a further approximation must be made to obtain secular dynamics, in which the populations and coherences of the system density matrix are independent. This approximation is justified when coherences oscillate quickly in comparison to the timescale of population dynamics, leading to their effects averaging out.\cite{breuer_petruccione} Comparisons with non-secular and numerically exact quantum calculations to confirm that the Born-Markov and secular approximations were appropriate, are found in the SI.

The population dynamics from the quantum master equations are assembled into a finite time Markov state model for QTPT. The transition rates between eigenstates in QTPT are equivalent to a jump process, which provides a physical interpretation for the treatment of the eigenstates as distinct elements in a Markov process. The dynamics modeled are those that would be observed in the case that the energy of the bath in contact with the system was continuously monitored.\cite{breuer_petruccione,jeske_cole,vogt_j_c,wiseman_milburn,garrahan_guta_cj} Within the quantum master equation, the tensor element describing the contribution of eigenstate $j$ to the change in population of eigenstate $i$ is\cite{balzer_stock} 
\begin{equation}
\begin{split}
    D_{iijj} &=  |R_{i,j}|^2 \int_0^\infty dt\, e^{-i \omega_{i,j} t} \langle B(0) B(t) \rangle_B  \\ &+|R_{i,j}|^2 \int_0^\infty dt\, e^{-i \omega_{j,i} t} \langle B(t) B(0) \rangle_B ,
\end{split}
\end{equation}
where $\langle ... \rangle_B$ indicates the average of an operator over the bath degrees of freedom in equilibrium and $|R_{i,j}|^2 = |\langle i| R | j \rangle|^2$. The states $| j \rangle$ and energies $E_i$ used to calculate frequencies, $\omega_{i,j}=(E_i-E_j)/\hbar$, correspond to the eigenstates and eigenvalues of $H_s$ in the absence of coupling to the bath. The rate of population transfer depends on the system coupling operator element and the one-sided Fourier transform of the bath correlation functions.
The population dynamics define the transition matrix with elements
\begin{equation}
    T_{ij} = (e^{\tau D}\sigma_{ii})_{jj} ,
\end{equation}
meaning population $j$ following propagation under the operator $D$ for time $\tau$, taken small, given the system was initialized in $\sigma_{ii}$, a density matrix in which all population is in energy eigenstate $i$.

The above formulation of the quantum master equation can be used to visualise the polariton energy landscape directly, by translating the equilibrium occupation probabilities and 
transition matrix into the equivalent relative free energies.\cite{WoodsKSSW23} 
Hence we obtain the free energy disconnectivity graph\cite{beckerk97,WalesMW98} in Fig. \ref{pes}d.
In this representation the vertical scale is the effective free energy, the bottom of each line corresponds to an eigenstate, and the eigenstates are connected together at a regular series of free energy thresholds when they can 
interconvert by any sequence of transition states that lies below the threshold.
Hence this construction provides a faithful account of the effective barriers and the organisation of the landscape.

The central quantity of transition path theory is the committor probability, $P_{b|a}(i)$, derived from the system of equations\cite{noe_schutte,metzner_schutte}
\begin{equation}
    P_{b|a}(i) - \sum_{j\epsilon I} T_{ij} P_{b|a}(j) = \sum_{j \epsilon b} T_{ij},
\end{equation}
which gives the probability for a system in eigenstate $i$ to visit eigenstate $b$ (the product state) before eigenstate $a$ (the reactant state) where $I$ is the set of all states that are neither $a$ nor $b$. Note that $P_{b|a}(b) = 1$ and $P_{b|a}(a)=0$.\cite{noe_schutte,metzner_schutte} Here we take the reactant and product states to be the lowest energy eigenstates localized on either side of the double well potential. Classically, the phenomenological transition state of a reaction is defined by a committor value of $1/2$.\cite{onsager1938initial,chodera2011splitting} However, for complex kinetic transition networks the productive paths and reactive visitation probabilities need to be considered together with committor values to diagnose the key dynamical bottlenecks.\cite{SharpeW21b,SharpeW21c} In QTPT we determine the pair of energy eigenstates where the probability changes from greater than to less than 1/2, defining a separatrix, as commitor eigenstates or transition eigenstates, whose role is analogous to a classical transition state in that they indicate a change of likely fate for the reactive pathway and are generally the bottle-neck states that limit the reactive flux.

\begin{figure}[t]
\begin{center}
\includegraphics[width=8.5cm]{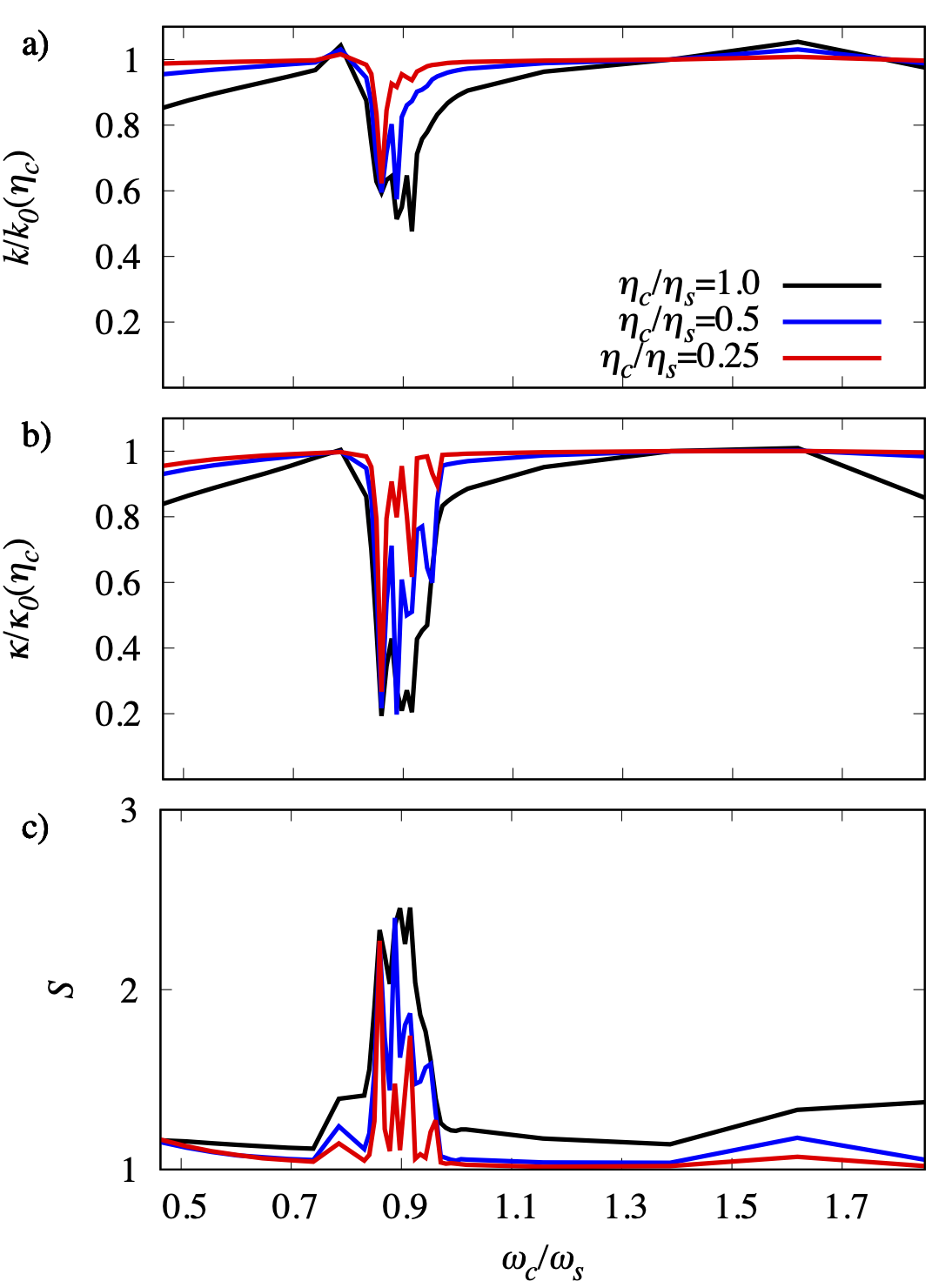}
\caption{
a) Rates of barrier crossing from eigenstate $1$ to $2$ at three light-matter coupling strengths as a function of $\omega_c/\omega_s$. See the Table 2 in the supporting information for values of $k_0(\eta_c)$, the reference values taken at $\omega_c/\omega_s=1.38$ for each coupling strength in order to plot them on the same axes.
 b) Transmission coefficient for the dominant reactive pathway as a function of $\omega_c/\omega_s$. See Table 2 in the SI for values of $\kappa_0(\eta_c)$, the reference values taken at $\omega_c/\omega_s=1.38$ for each coupling strength in order to plot them on the same axes.
 c) Path entropy for the three light-matter coupling strengths as a function of $\omega_c/\omega_s$.
}
\label{rates}
\end{center} 
\end{figure}
From the committors, we found the barrier crossing rate, $k$,\cite{anderson_schile_limmer,noe_schutte,metzner_schutte} as a function of $\omega_c/\omega_s$, where $\omega_s$ is the approximate harmonic frequency of the proton, to look for resonance rate modification effects. In Fig. \ref{rates} a), the barrier crossing rate relative to a reference $k_0(\eta_c)$ is provided for three different light-matter coupling strengths, $\eta_c$. The light matter coupling strength is defined as, \cite{li_mandal_2021}
\begin{equation}
    \eta_c=\frac{\partial \mu(R)}{\partial R}\Bigr|_{R_0}\sqrt{\frac{\hbar}{2\omega_s M }}\frac{\chi}{\hbar \omega_c},
\end{equation}
where $R_0$ is the equilibrium position of $R$ in the reactant well. The coupling strength is held constant by modifying $\chi$ in proportion to $\omega_c$. For convenience, in this work the default light-matter coupling is defined by $\eta_s$ where $\eta_s=0.02$ and all coupling strengths will be defined relative to this value. This coupling strength is similar to that addressed by Lindoy and coworkers in their recent work.\cite{lindoy_nature} It is much smaller than the coupling strengths often employed in classical theoretical treatments, however, it is still relatively very strong coupling compared to experiments. Note that we are considering a single molecule coupled to the cavity, and thus the relevant coupling strength is related to the polaritonic splitting by a factor dependent on the number of solutes.\cite{du_poh} 

Although the rates have been normalized for visual comparison, the absolute barrier crossing rates in Fig. \ref{rates} a) decrease with increasing $\eta_c$. This change can be explained by regarding the photon coordinate as an extra degree of freedom imposing friction on the proton coordinate and indicates the system is in the high friction limit. This interpretation is qualitatively consistent with previous classical theories;\cite{li_mandal_2021,yang_cao} it does not imply a high friction limit for the quantum bath and only explains relations between coupling strength and rate for the two degrees of freedom explicitly modeled in the system. However, we find a clear resonance rate suppression near $\omega_c/\omega_s=0.9$. The observed resonant rate suppression does not occur exactly at $\omega_c/\omega_s=1$ because the system is anharmonic and the energy gap between proton vibrational states prior to coupling to the photon coordinate is below $\omega_s$ for the higher energy states involved in barrier crossing reactions. Higher $\eta_c$ values result in stronger resonances with multiple peaks.

To identify reactive pathways and glean mechanistic insight into the rate suppression, we first calculated the reactive flux between eigenstates. In a detailed balance system, the reactive flux between any two eigenstates for the reaction $a \rightarrow b$ is given by
\begin{equation}
    f_{i,j}^{a,b} = \pi_i P_{a|b}(i) T_{i,j} P_{b|a}(j) \;\;\; i \ne j ,
\end{equation}
where $\pi_i$ is the equilibrium population of $i$ and $P_{a|b}(i) = 1 - P_{b|a}(i)$. 
The net fluxes between eigenstates are treated as edge weights in a graph with all of the eigenstates as vertices. The maximum flux pathway between $a$ and $b$ is then extracted with Dijkstra's algorithm.\cite{metzner_schutte,dijkstra} This procedure is repeated to obtain a reactive path ensemble. The committor eigenstates are defined as the last state along a reactive pathway with $P_{b|a}(i) < 1/2$ and the first state along a reactive pathway with $P_{b|a} > 1/2$. These states characterize the system immediately before and immediately after it has committed itself to completing the reaction. 

We inspected the dominant barrier crossing pathways extracted by QTPT as a function of photon frequency and studied their effectiveness in the vicinity of resonant rate suppression. A useful decomposition of the rate is given by defining $\kappa = k_{1}/\exp[-\beta \Delta E^\dagger]$, where $k_{1}$ is the rate associated with the dominant reactive path and $\Delta E^\dagger$ is the activation energy, computed as a difference between the ground state energy and the largest energy visited on the dominant path. This is smaller than the classical barrier height due to zero point energy and tunneling through the barrier. 

The decrease in $\kappa$ in Fig. \ref{rates} b) corresponds closely to the observed resonant fall in rates in Fig. \ref{rates} a) indicating that a lack of transmission rather than a change in activation energy in the dominant pathway is at least partially responsible for the resonant rate decrease. Indeed, for the range of $\omega_c$ considered, the activation energy of the dominant path varies rather little. This behavior is consistent with observations that the change in the barrier height due to coupling to the cavity is not the primary mechanism for rate suppression.\cite{thomas_george_shalabney} At the frequencies where we observe a suppression in the rate, we  also find that the number of reactive pathways participating in the reaction increases. This effect is quantified by the path entropy,
$     S=-\sum_{\alpha} \hat{f_\alpha} \ln(\hat{f_ \alpha})$ ,
where $f_{\alpha}= \min_{i,j} [f_{i,j}^{a,b}]$ for $i$ and $j$ along unique reactive paths and $\hat{f_\alpha} = f_\alpha/\sum_\alpha f_\alpha$. The spikes of $S$ seen in Fig. \ref{rates} c) correspond with the resonant rate decreases, revealing that a larger variety of reactive pathways contribute to the ensemble under resonance conditions. This result indicates that the dominant pathway is being rendered less effective and other pathways are more important.

\begin{figure}[t]
\begin{center}
\includegraphics[width=8.5cm]{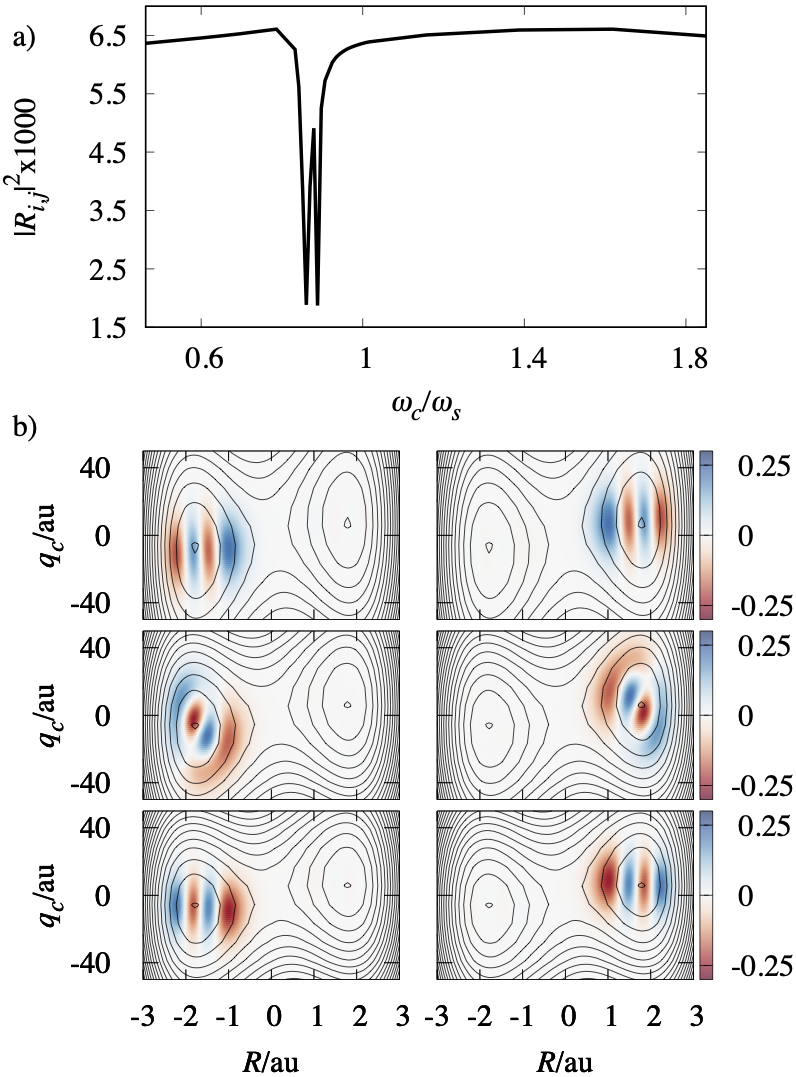}
\caption{a) Overlap element squared for the committor eigenstates of the dominant pathway as a function of $\omega_c/\omega_s$ for the case where $\eta_c/\eta_s=0.5$. b) Spatial distribution of the real part of the pre-committor (left) and post-committor (right) eigenstate wavefunctions for $\eta_c/\eta_s=1.0$ at $\omega_c/\omega_s=0.74$ (top), $\omega_c/\omega_s=0.916$ (middle) and $\omega_c/\omega_s=1.02$ (bottom) with energy contours spaced 0.01 au apart. 
}
\label{overlap}
\end{center} 
\end{figure}

The origin of the resonant effect that decreases the effectiveness of the dominant pathway is apparent when we inspect the jump between the pre and post-committor eigenstates for the dominant pathway under resonance conditions. The dominant pathway, in all cases, is a tunneling pathway as both pre and post-committor eigenstates have energies below the potential barrier. The square coupling operator elements, $|R_{i,j}|^2$, linking the committor eigenstates of the dominant pathway for $\eta_c/\eta_s = 0.5$ in Fig. \ref{overlap} a), are directly proportional to inter-eigenstate transfer rates in the quantum master equations, and show the same double-peaked pattern of resonance suppression in the overall rates observed for this coupling strength in Fig.~\ref{rates} a).  
Poor overlap at the committor jump results in resonant rate suppression. The resonance effect observed in $|R_{i,j}|^2$ is stronger than that observed in the rates themselves, but this contribution to the rate is offset by the modest increase in other reactive pathways in the system which contribute to the rate in larger amounts on resonance.

Interrogation of the spatial distribution of the committor eigenstate wavefunctions in Fig. \ref{overlap} b) explains the source of the poor overlap for the dominant path under resonance conditions. At $\omega_c/\omega_s=0.74$ and $\omega_c/\omega_s=1.02$, below or above resonance, the wavefunctions of the pre and post-committor eigenstates closely resemble conventional harmonic oscillator wavefunctions. However, on resonance at $\omega_c/\omega_s=0.916$, the committor wavefunctions do not resemble harmonic oscillator wavefunctions, instead exhibiting mode hybridization. These polaritonic wavefunctions appear to be rotated relative to the coordinate $R$ by which the system is coupled to the bath, explaining the poor overlap.

Resonance with the cavity resulting in the formation of polaritonic states along the critical reactive pathways in the barrier crossing reactions results in a sharp decrease in barrier crossing rates. This phenomenon agrees with the sharp resonance effects observed experimentally\cite{thomas_george_shalabney,lather_bhatt_thomas,thomas_lethuillier-karl} and observed by Lindoy and coworkers using fully quantum dynamical simulations.\cite{lindoy_nature} The simulations by Lindoy and coworkers indicated sharp rate increases or decreases depending on their choice of cavity loss and bath structure, an effect they similarly attributed to changes in bath interactions upon light-matter hybridization. In contrast to this work, Lindoy and coworkers did not observe rate modifications for lossless cavities. However, we note that Lindoy and coworkers employed models mostly in the energy diffusion regime. As the observed resonant suppression effect depends on the dominance of deep-tunneling mechanisms, which depend on a weakly coupled environment and many well-localized energy eigenstates below the barrier, it is unsurprising that dynamics in the lossy cavity of Lindoy and coworkers differ from those we have observed.

The resonant suppression observed here is a fundamentally open quantum system phenomenon that will disappear in the high temperature, classical limit, where tunneling mechanisms do not play a role, or when the cavity coupling becomes sufficiently weak as to no longer form a polariton with a single molecule. Any theory that is unable to explicitly account for the hybridization of light-matter states to form polaritons, or is unable to account for the interactions of the bath with polaritonic states consistently, will not reproduce these resonant effects. This observation explains the failure of Grote-Hynes theory to uncover the sharp resonance effects\cite{galego_climent,cga_jorge_yz,zhadanov,li_nitzan_subotnik}, even though it correctly identifies the origin of suppression being an altered reactive flux, rather than a change in activation energy. We suggest that further studies into the potential of polaritonic effects, in selective bond-breaking reactions and other applications, should make use of methods that explicitly address the formation of polaritons and their interaction with environmental fluctuations.

{\bf Supporting Information Available.} Detailed description of the model system, specification of parameters, additional disconnectivity graphs and background, mean first passage time comparisons on and off resonance, Redfield theory background, comparison of secular Redfield, non-secular Redfield and numerically exact quantum calculations.

{\bf Data Availability.} Data and code supporting this study are available in Zenodo at 10.5281/zenodo.8002313.

{\bf Acknowledgments}. We would like to thank David Manolopoulos for useful discussions. M.C.A., T. P. Fay and D.T.L. were supported by the U.S. Department
of Energy, Office of Science, Basic Energy Sciences, CPIMS Program Early Career Research Program under Award No. DE-FOA0002019. E. J. Woods gratefully acknowledges support from the Engineering and Physical Sciences Research Council [grant numbers EP/R513180/1, EP/N509620/1]. D. J. Wales also gratefully acknowledges support from the Engineering and Physical Sciences Research Council [grant number EP/N035003/1].

\section*{References}
\bibliographystyle{achemso}
\bibliography{ref}
\end{document}


\maketitle

\section{System Setup}

The full Hamiltonian of the open system is $H = H_s + H_B + R \otimes B$ where $B=\sum_k c_k R_k$. Operator $R$ is the proton position operator and $R_k$ are positions of harmonic bath modes with the parameter $c_k$ a coupling strength parameter. The value of $\hbar$ is taken as $1$ throughout. The system Pauli-Fierz\cite{pauli_fierz} Hamiltonian describing a Shin-Metiu\cite{shin_metiu} model has proton mode ($R$) and photon mode ($q_c$) giving,
\begin{equation}
    H_s = P^2/(2M) + U(R) + p_c^2/2 + \omega_c^2/2 \left(
q_c + \sqrt{2/(\hbar \omega_c^3}) \chi \mu(R)
\right)^2
\end{equation}
where $p_c$ and $q_c$ are photon momentum and position operators, $M$ is the proton mass, $R$ and $P$ are the proton position and momentum operators, $\omega_c$ is the photon frequency, $\chi$ is a coupling strength parameter, the dipole operator is
\begin{equation}
    \mu(R) = -1.90249\;\mathrm{tanh}(1.26426R) + 0.37044R,
\end{equation}
the potential energy surface is
\begin{equation}
U(R) = -0.021088R^2 + 0.0033108R^4 + 0.033161 + 3.6749\times10^{-6}R + H_{ls}(R),
\end{equation}
and the negligibly small renormalization of the system energy levels imposed by the bath is
\begin{equation}
    H_{ls} = \omega_b \eta R^2/(M\pi),
\end{equation}
with $\omega_b$ representing the cutoff frequency for the bath oscillators in the spectral density description and $\eta$ being the system-bath coupling strength.
The dipole is assumed to align perfectly along $R$. The bath Hamiltonian,
\begin{equation}
H_{B} = \sum_k P_k^2/2 +  \omega_k^2 R_k^2/2,
\end{equation}
is an unstructured, infinite set of harmonic oscillators with momenta $P_k$, position $R_k$, frequency $\omega_k$, and $\beta = 1/(k_B T)$ describing the temperature, T, where $k_B$ is Boltzman's constant. The coupling operator links the system and bath modes along their position coordinates. The spectral density describing the distribution of the harmonic oscillators in the bath is 
\begin{equation}
    J(\omega) = \pi/2 \sum_k \frac{c_k^2}{ \omega_k} \delta(\omega - \omega_k) = \eta \omega e^{-|\omega|/\omega_b},
\end{equation} 
an Ohmic exponential form.

\renewcommand{\arraystretch}{1.5}
\begin{table}
\centering
\begin{tabular}{|p{2.0cm} | p{7.5cm}| }
\hline 
\hline 
Parameter  & value  (atomic units unless specified)  \\
\hline

$\beta$ & 1052.584413  \\
$\omega_s$ & 0.00677687  \\
$\omega_b$ & 0.006269431  \\
$M$ & 1836  \\
$\chi_0$ $^*$  & 0.002535471   \\
$\eta$ & 0.0018228  (unitless)\\
\hline
\end{tabular}

\begin{tablenotes}
   \centering
   \item[*] $^*$Note that $\chi_0$ is the value of $\chi$ when $\omega_c/\omega_s = 1$ and $\eta_c/\eta_s=1$.
  \end{tablenotes}
\caption{Parameters employed during simulation of the polariton model} 
\label{tab1}
\end{table}

\renewcommand{\arraystretch}{1.5}
\begin{table}
\centering
\begin{tabular}{|p{2.0cm} | p{3.5cm}| p{3.5cm} | }
\hline 
\hline 
$\eta_c$  & $k_0(\eta_c)/{\mathrm{au}}\;^{-1}$ & $\kappa_0(\eta_c)/\;{\mathrm {au}}^{-1}$ \\
\hline

$\eta_s$ & $7.643\times10^{-17}$ & $1.667\times10^{-8}$\\
$\eta_s/2$ & $9.782\times10^{-17}$  &  $2.180\times10^{-8}$\\
$\eta_s/4$ & $1.042\times10^{-16}$ & $2.331\times10^{-8}$\\
\hline
\end{tabular}
\caption{Parameters employed in rate and transmission constant specifications} 
\label{tab2}
\end{table}

Our results are sensitive to the magnitude of the applied linear bias that breaks the symmetry of the double well system. A bias that is too large can fundamentally alter the system and change the eigenstructure and hence locations of resonance between the photon and proton mode, whereas an insufficient linear bias will not sufficiently break the symmetry. The bias was selected to be as small as possible. Doubling or halving the linear bias did not change the trend in the observed resonant behavior, although it did change absolute rates and relative magnitude of the resonant rate decrease. Decreasing the linear bias further resulted in lack of localization of relevant eigenstates.

To check that the imposed linear bias had not changed the structure of the system in an unexpected way, we prepared disconnectivity graphs\cite{BeckerK97,walesmw98} for several relative frequencies and light matter coupling strengths.\cite{schutz_schachenmayer}

To obtain the free energy disconnectivity graphs, we translate the equilibrium probabilities of each eigenstate $\pi_a$, and the transition rates $D_{bbaa}$, to effective free energies of the eigenstates $g_{a}$, and the transition states connecting them, $g^{\dagger}_{ab}$,
\begin{equation}
    \tilde{g}_a(T) = -k_\mathrm{B}T\ln(\pi_a),
\end{equation}
\begin{equation}
    g^{\dagger}_{ab}(T) = g_{a}(T) - k_\mathrm{B} T\ln(D_{bbaa}) + k_\mathrm{B}T\ln(k_\mathrm{B}T/h),
\end{equation}
where $h$ is the Planck constant, and the subscript $ab$ refers to a transfer from eigenstate $a$ to $b$.\cite{WoodsKSSW23}
Disconnectivity graphs are a visualization tool that illustrate how eigenstates are connected. The branching structure of the graphs reflect the structure of the potential energy surface, in this case the double well potential. The double funnel structure observed here is well known for molecular systems that feature competition between alternative morphologies.\cite{walesmw98,MillerDW99a,DoyeMW99a,WalesB06} 

\begin{figure}[t]
\begin{center}
    \includegraphics[width=\textwidth]{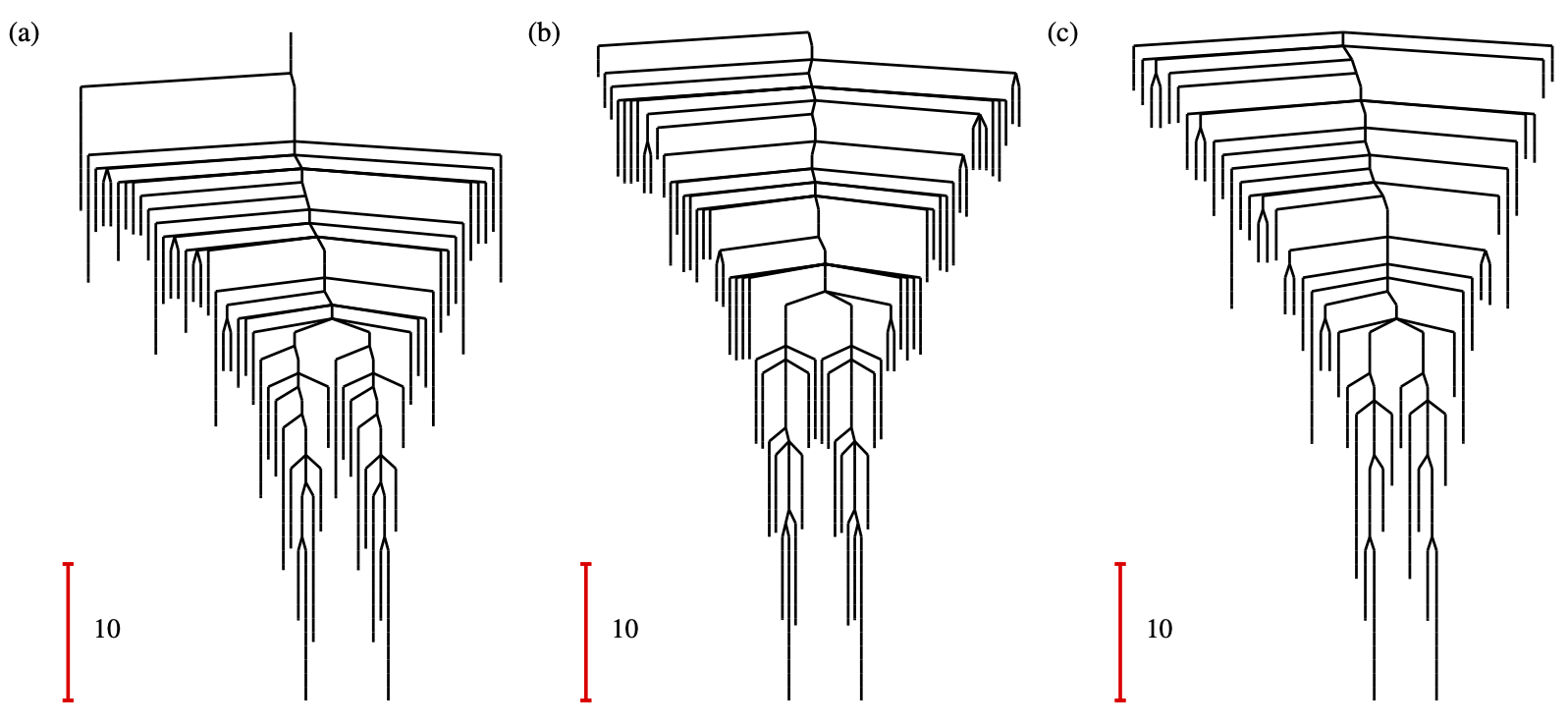}
    \caption{Disconnectivity graphs for $\eta_c/\eta_s=1$ light-matter coupling strength with a) $\omega_c/\omega_s=0.74$, b) $\omega_c/\omega_s=0.916$ c) $\omega_c/\omega_s=1.39$. The vertical axis corresponds to free energy in units of $k_\mathrm{B}T$, with a scale bar of $10k_\mathrm{B}T$.  The branches ending at each eigenstate are positioned on the horizontal axis to highlight the organisation of the landscape. }
    \label{tree}
\end{center}
\end{figure}

Disconnectivity graphs generated below and above resonance are displayed in Fig.~\ref{tree} a) and c), with the graph for on-resonance in \ref{tree} b). The symmetric double well is evident in the symmetric split of the branches at the bottoms of both disconnectivity graphs, indicating that the linear bias has not appreciably altered the fundamental structure of the system. The higher free energy placement of the transition state linking eigenstates $1$ and $2$ for the on resonance case in Fig. \ref{tree} b) is  small but still significant, indicating a decrease in transition rate.

The observed resonant behavior was not sensitive either to the precise form of the bath spectral density or to the bath cutoff or coupling strength. Resonant suppression was observed with $\eta$ and $\omega_c$ both increased by a factor of $10$, and when a Debye\cite{nitzan} rather than Ohmic exponential bath was implemented with $\eta$ selected so as to keep the absolute values of the rates approximately equal when off resonance. 

Provided that the barrier height remained large in comparison to the thermal energy, $k_\mathrm{B} T$, the observed resonant rate suppression was preserved. Doubling the temperature did not eliminate the observed resonance. However, increasing the temperature by a factor of ten resulted in a classical limit, in which a great deal of density was not concentrated in the lowest energy eigenstates in either well and the majority of reactive pathways passed above the barrier, eliminating resonance effects that are fundamentally characteristic of quantum tunneling.

\begin{figure}[t]
\begin{center}
\includegraphics[width=\textwidth]{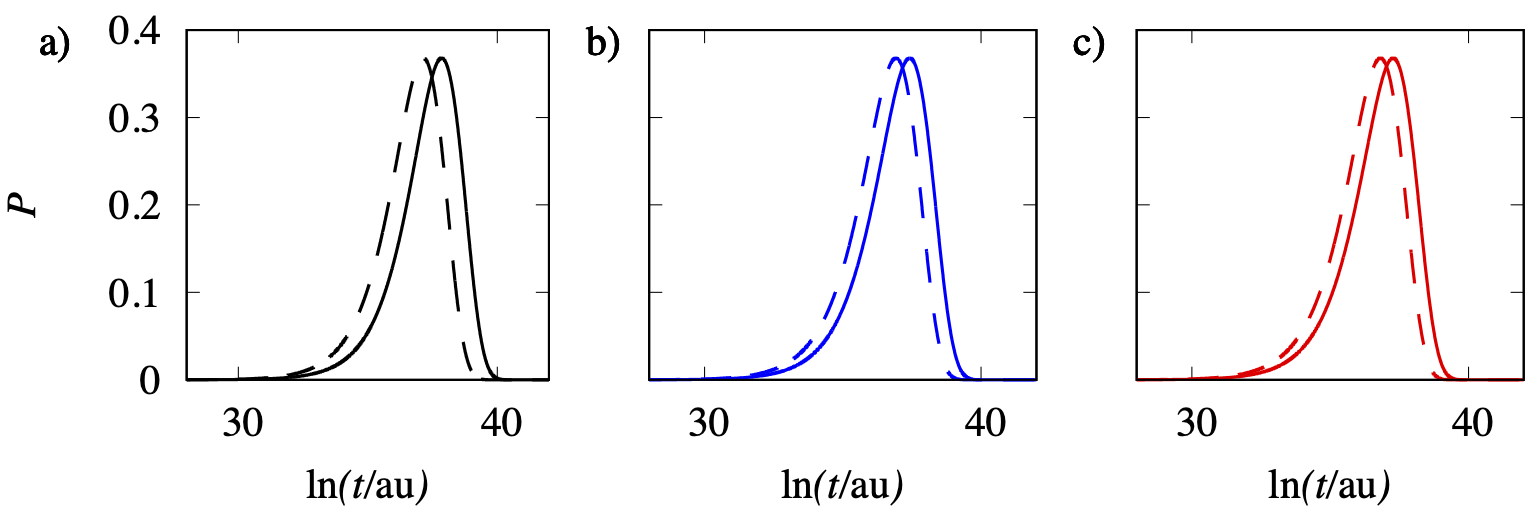}
\caption{First passage time distributions for passage from the lowest to the second lowest energy eigenstate, a) $\eta_c/\eta_0 = 1$ b) $\eta_c/\eta_0 = 0.5$ c) $\eta_c/\eta_0 = 0.25$, showing larger passage times for on resonance systems. Dashed lines indicate off-resonance and solid lines indicate on-resonance.
}
\label{fpt}
\end{center}
\end{figure}

We calculated the first passage time distributions\cite{wales_2022,swinburne_wales} at several key resonance and off-resonance $\omega_c$ values at the three different coupling strengths in \ref{fpt}. We start from the master equation for the eigenstate populations,
\begin{equation}
    \frac{\partial \boldsymbol{\sigma}}{ \partial t}= \bf{D} \boldsymbol{\sigma},
\end{equation}
with rate matrix $\bf{D}$ formed from the block diagonal elements of the Redfield tensor, $(\mathbf{D})_{ij}=D_{jjii}$. We decompose the matrix $\bf{D}=\bf{K}-\bf{A}$ into a matrix $\bf{K}$ with elements $K_{ij}$ representing the transition rate from $i$ to $j$,  and a diagonal rate matrix $\bf{A}$ of total escape rates, $A_{ij}=\delta_{ij} \sum_{\gamma \neq j} K_{j\gamma}$. The time dependent population vector $\boldsymbol{\sigma}$ is formed from the diagonal of the density matrix in the eigenstate representation, $(\boldsymbol{\sigma})_i=\sigma_{ii}$. To compute the first passage time distribution between eigenstates $a$ and $b$, we work in the reduced state space $\mathcal{S}$, which contains all states, apart from the product state $b$. The transition matrices are reduced to $\bf{D}_{\mathcal{S}}$ and $\bf{K}_{\mathcal{S}}$, and only contain the transitions between states in $\mathcal{S}$. The escape rates to $b$ are only retained in the diagonal entries of ${\bf A}_{\mathcal S}$, which is the corresponding subset of $\bf{A}$. By writing ${\bf D}_{\mathcal{S}}$ in terms of its eigenvalues $-\lambda_{\ell}$ and its left and right eigenvectors ${\bf w}_{\ell}^{L}$ and ${\bf w}_{\ell}^{R}$, we can write the first passage time distribution from eigenstates $a$ to $b$ as
\begin{equation}
    p(\theta)=\sum_{\ell}\lambda_{\ell} e^{-\lambda_{\ell} \theta}  {\bf 1}
        \left({\bf w}^R_{\ell} \otimes
        {\bf w}^L_{\ell}\right) \boldsymbol{\sigma}_{\mathcal{S}}(0),
\end{equation}
where ${\bf 1}$ is a row vector of ones and $\boldsymbol{\sigma}_{\mathcal{S}}(0)$ is the initial eigenstate occupation probability. To aid with visualization we transform to a probability distribution in $y=\ln(\theta)$,\cite{wales_2022}
\begin{equation}
    \mathcal{P}(y)=\sum_{\ell=1}\lambda_{\ell} e^{y-\lambda_{\ell} e^y}  {\bf 1}
        \left({\bf w}^R_{\ell} \otimes
        {\bf w}^L_{\ell}\right) {\boldsymbol{\sigma}}_{\mathcal{S}}(0).
\end{equation}
Our open source program PATHSAMPLE,\cite{PATHSAMPLE} was used to perform the first passage time calculations, and the results are shown in Fig. \ref{fpt}. 
The first passage time distribution is shifted to longer times at resonance, indicating that all barrier crossing reactions are slowed considerably under resonance effects.

\section{Secular Redfield Theory}

Secular Redfield theory was employed to describe the dynamics of the system. \cite{balzer_stock,breuer_petruccione,nitzan}
The Redfield tensor,
\begin{equation}
    D_{ijkl} = \Gamma^+_{ljik} + \Gamma^{-}_{ljik} - \delta_{lj}\sum_m \Gamma^+_{immk} - \delta_{ik}\sum_m \Gamma^{-}_{lmmj}
\end{equation}
is formed from elements of the system component of the coupling operator and one-sided Fourier transforms of the bath correlation functions giving
\begin{equation}
    \Gamma^+_{ljik}=R_{lj}R_{ik} \int_0^\infty dt e^{-i \omega_{i,k}t} \langle B(0) B(t) \rangle
\end{equation}
and 
\begin{equation}
    \Gamma^-_{ljik}=R_{lj}R_{ik} \int_0^\infty dt e^{-i \omega_{l,j}t} \langle B(t) B(0) \rangle
\end{equation}
with components of $R$ defined as
\begin{equation}
    R_{lj} = \langle l | R | j \rangle.
\end{equation}
In secular Redfield theory, the change of population of the energy eigenstates of the density matrix is described as,
\begin{equation}
    \frac{\partial \sigma_{ii}(t)}{\partial t}  = \sum_j D_{iijj} \sigma_{jj}(t)
\end{equation}
whereas the coherences,
\begin{equation}
    \sigma_{ij}(t) = \sigma_{ij}(0) e^{(-i \omega_{ij} + D_{ijij}) t},
\end{equation}
are completely decoupled from the populations and undergo exponential decay. 
Provided the initial condition, $\sigma_0$, is incoherent, no coherences will exist in the system evolution. This is the fundamental justification for QTPT, as it allows us to conceptualize reactive pathways through a quantum system as a series of jumps between eigenstates where the probability of the jump depends only on the populations of other eigenstates.

The photon coordinate is described by a harmonic oscillator basis of dimension 70. The proton coordinate is described by a Colbert-Miller DVR basis\cite{colbert_miller} with dimension of 101 and $\delta=0.1 \; \mathrm{au}$. The basis was trimmed following diagonalization and secular Redfield propagation was carried out on the lowest 200 energy eigenstates only.

\section{QTPT and Simulation Details}
Committor probabilities provide reactive fluxes between each eigenstate. From the reactive fluxes, the full rate of reaction,
\begin{equation}
    F = \sum_{ j \ne a} f_{a,j}^{a,b} = \sum_{j \ne b} f_{j,b}^{a,b},
\end{equation}
is found by summing all reactive flux leaving $a$ or entering $b$. The reaction rate in a detailed balance system,
\begin{equation}
k=\frac{F}{\tau \Pi_a},
\end{equation}
is found by dividing the flux by $\tau$, the time over which the MSM was generated, and the probability for the system to be moving forward from $a$ to $b$, $\Pi_a=\sum_i \pi_i P_{a|b}(i)$.\cite{noe_schutte,metzner_schutte} The Markov state models were generated from secular Redfield\cite{balzer_stock,redfield,nitzan} computations beginning with all density in each eigenstate in turn and lasting for $2500 \; \mathrm{au}$ with timesteps of $5 \; \mathrm{au}$. Committors ($P_{b|a}(i)$) and reverse committors ($P_{a|b}(i)$) were solved independently and then detailed balance was imposed such that $P_{a|b}(i) = 1 - P_{b|a}(i)$ with the smaller of the values used to correct the larger value. The code to perform these calculations and accompanying data is available in a repository.\cite{tpt_ci}

\section{Comparison With Non-secular Dynamics}

\begin{figure}[t]
\begin{center}
\includegraphics[width=11.5cm]{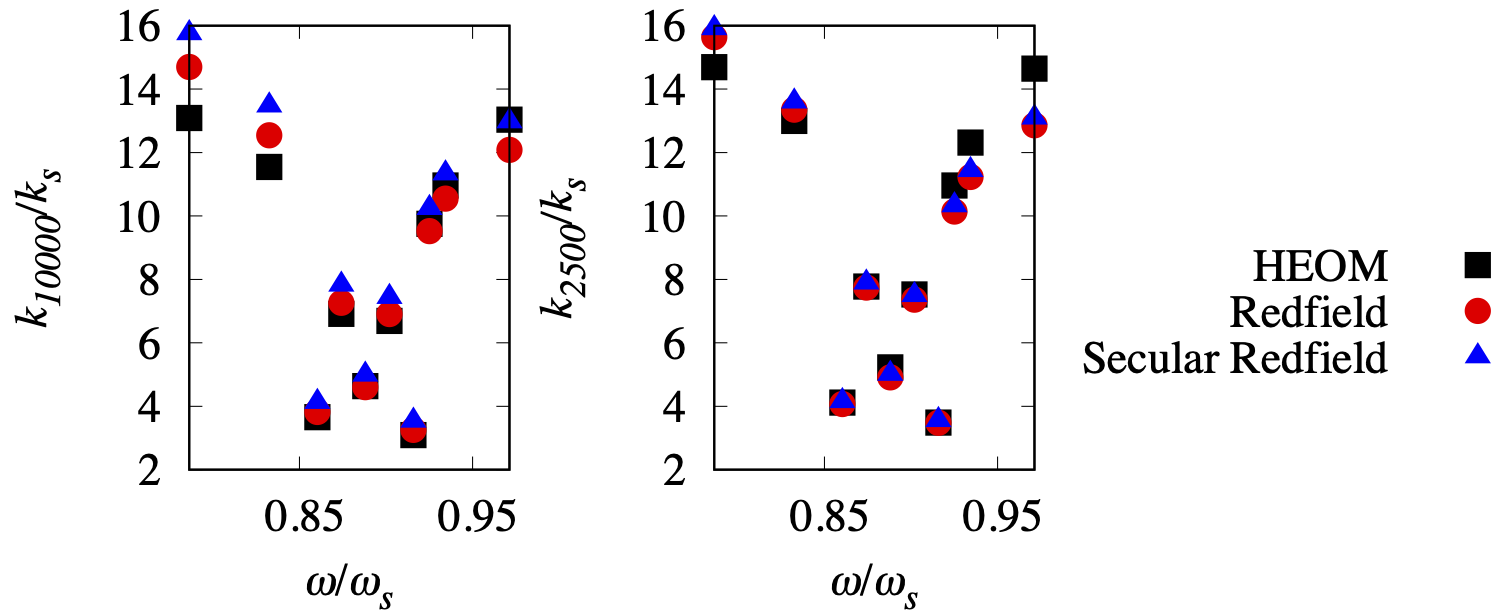}
\caption{Comparison of rate estimates from hierarchical equation of motion, Redfield and secular Redfield calculations for the committor jumps as a function of $\omega_c/\omega_s$ around the resonance for $\eta_c/\eta_s=1.0$ with a Debye bath, $k_s=1*10^{-9}$, $k_{10000}$ indicating the estimate from a 10000 au length calculation of each method and $k_{2500}$ indicating the estimate from a 2500 au length calculation for each method.
}
\label{heom_compare}
\end{center}
\end{figure}

To validate the performance of secular Redfield\cite{balzer_stock} theory, which depends on several strong assumptions which can be in question in a system with near-degeneracies in the energy eigenspectrum,\cite{breuer_petruccione,egorova_kuhl} we estimated and compared the rates of population transfer for the committor jumps around the resonance for Redfield,\cite{redfield,nitzan} secular Redfield and the formally exact hierarchical equations of motion (HEOM) method using the [N/N] Padé decomposition of the bath correlation functions with N=30 and two layers (L=2).\cite{tanimura_kubo,fay_2022,heom-lab,Hu2011} This treatment required truncation to 50 eigenstates and modification of the form of the spectral density to a Debye form,\cite{nitzan} 
\begin{equation}
    J(\omega) = \pi/2 \sum_k \frac{c_k^2}{M_k \omega_k} \delta(\omega - \omega_k)  = \eta \omega/\left( \omega^2 + \omega_b^2\right),
\end{equation}
 with $\omega_b$ given in Table \ref{tab1} and $\eta=6.601876175*10^{-8}$, which maintained similar off resonance rates to those observed under the Ohmic exponential bath description. In cases where the secular approximation is not applicable, it may overestimate tunneling rates between states with similar energies, a phenomenon which would be immediately apparent in short-time propagation comparison between the three methods. Secular Redfield theory is known to overestimate tunneling rates\cite{ishizaki_fleming} in comparison to Redfield and formally exact HEOM, however, rates estimated from cumulative population transfed, meaning the rate estimate is the population following propagation divided by the propagation time, for these critical, rate-limiting jumps were in very good agreement at $2500 \; \mathrm{au}$, the time used to generate the Markov matrices for QTPT, and at $10000 \; \mathrm{au}$, as demonstrated in Fig.~\ref{heom_compare}, indicating that the secular approximation is applicable to this system.

\bibliography{ref}